\documentclass{aa}

\usepackage{graphicx}
\usepackage{txfonts}
\usepackage{color}
\begin{document}

   \title{Search for exoplanet around northern circumpolar stars\thanks{Based on observations made with the BOES instrument on the 1.8~m telescope at Bohyunsan Optical Astronomy Observatory in Korea.}}

   \subtitle{Four planets around HD 11755, HD 12648, HD 24064, and 8~Ursae Minoris}

   \author{B.$-$C. Lee\inst{1,2},
          M.$-$G. Park\inst{3},
          S.$-$M. Lee\inst{1},
          G. Jeong\inst{1,2},
          H.$-$I. Oh\inst{3},
          I. Han\inst{1},
          J. W. Lee\inst{1},
          C.-U. Lee\inst{1},
          S.-L. Kim\inst{1},
          \and
          K.$-$M. Kim \inst{1}
          }

   \institute{Korea Astronomy and Space Science Institute, 776,
		Daedeokdae-Ro, Youseong-Gu, Daejeon 305-348, Korea\\
	      \email{[bclee;smlee;tlotv;iwhan;jwlee;leecu;slkim;kmkim]@kasi.re.kr}
	    \and
		Korea University of Science and Technology,
        Gajeong-ro Yuseong-gu, Daejeon 305-333, Korea
	    \and
	     Department of Astronomy and Atmospheric Sciences,
	     Kyungpook National University, Daegu 702-701, Korea\\
	      \email{[mgp;ymy501]@knu.ac.kr}
             }

   \date{Received July xx, 2015; accepted xxx xx, 2015}


  \abstract
   {}
   {This program originated as the north pole region extension of the established exoplanet survey using 1.8 m telescope at Bohyunsan Optical Astronomy Observatory (BOAO). The aim of our paper is to find exoplanets in northern circumpolar stars with a precise radial velocity (RV)
survey.}
   {We have selected about 200 northern circumpolar stars with the following criteria: $\delta$ $\geq$ 70\,$^{\circ}$, 0.6 < \emph{B$-$V} < 1.6, \emph{HIPPARCOS}$_{\rm scat}$ < 0.05 magnitude, and 5.0 < \emph{m}$_{v}$ < 7.0. The high-resolution, fiber-fed Bohyunsan Observatory Echelle Spectrograph (BOES) was used for the RV survey. Chromospheric activities, the \emph{HIPPARCOS} photometry, and line bisectors were analyzed to exclude other causes for the RV variations.}
   {In 2010, we started to monitor the candidates and have completed initial screening for all stars for the last five years.
   We present the detection of four new exoplanets. Stars HD 11755, HD 12648, HD 24064, and 8 UMi all show evidence for giant planets in Keplerian motion.
   The companion to HD 11755 has a minimum mass of 6.5 $\it M_\mathrm{Jup}$ in a 433-day orbit with an eccentricity of 0.19. HD 12648 is orbited by a companion of minimum mass of 2.9 $\it M_\mathrm{Jup}$ having a period of 133 days and an eccentricity of 0.04. Weak surface activity was suspected in HD 24064. However, no evidence was found to be associated with the RV variations. Its companion has a minimum mass of 9.4 $\it M_\mathrm{Jup}$, a period of 535 days, and an eccentricity of 0.35. Finally, 8 UMi has a minimum mass of 1.5 $\it M_\mathrm{Jup}$, a period of 93 days with an eccentricity of 0.06.
   }
   {}

   \keywords{stars: individual: HD 11755, HD 12648, HD 24064, 8 Ursae Majoris (HD 133086) --- stars: planetary systems  ---  techniques: radial velocities
   }

   \authorrunning{B.-C. Lee et al.}
   \titlerunning{Search for exoplanet around northern circumpolar stars}
   \maketitle
%

\section{Introduction}
Since the 1990s, exoplanet surveys by precise radial velocity (RV) method have been conducted in many observatories around the world.
Nearly 25 years of concerted RV monitoring has revealed an interesting and diverse planetary systems.
Almost all initial exoplanet search programs have focused on G, K main sequence (MS) as host stars because they are bright enough to have a high signal-to-noise (S/N) ratio and have an ample number of spectral lines for precise RV measurements.
However, due to deficit of candidate stars in this stage for ground-based observations, searches have been expanded gradually to evolved stars (Frink et al. 2002; Setiawan et al. 2003; Sato et al. 2003; Hatzes et al. 2005; D{\"o}llinger et al. 2007; Han et al. 2010).
There is no doubt that most planet search programs reflect bias towards relatively bright stars and against fainter M dwarfs. However, M dwarfs are the most numerous class of stars in the solar neighborhood, comprising $\sim$ 75\% of the stars (Reid et al. 2002). Low-mass stars may be the most typical targets of planet formation (Lada 2006) and were collectively monitored by several groups (Endl et al. 2003; Wright et al. 2004; Bonfils et al. 2005). Now M dwarfs are the most active observational target (Anglada-Escud{\'e} et al. 2012; Bonfils et al. 2013; Howard et al. 2014; Bowler et al. 2015).

Most of the ground-based RV surveys have been conducted in the form of an all-sky survey in accordance with the classification of spectral type and luminosity class and developed to maximize the RV accuracy of a star. But, this strategy causes a bias.
Most of RV observatories in the northern hemisphere are focusing observations on specific spectral or luminosity class rather than targeting a specific area. Due to the geographical constraints and the effective operation of the telescope, high declination zone has not been observed actively. Even in Hawaii where large telescopes for spectroscopic observatories are densely populated, observations of more than 70\,$^{\circ}$ declination are not effective.
As a result, such area has become a likely blind spot of planet survey
(+70\,$^{\circ}$  < $\delta$ < +90\,$^{\circ}$  and $-$70\,$^{\circ}$ < $\delta$ < $-$90\,$^{\circ}$ ). Around 20  exoplanets have been found in this region (6\% of the total area) corresponding to $\sim$3\% of detection by the RV method and $\sim$1\% of the total exoplanet detection.
Here, we introduce a new exoplanet survey, a “\textbf{S}earch for \textbf{E}xoplanet around \textbf{N}orthern circumpolar \textbf{S}tars (SENS)”, which uses the 1.8~m telescope at Bohyunsan Optical Astronomy Observatory (BOAO) in Korea to search for planets among several survey projects.

In this paper, we present the first result from the survey. In Sect. 2, we introduce the SENS program. Sect. 3 describes the observational data and reduction procedures. We detail the stellar characteristics in Sect. 4 and the RV analysis and the nature of the RV measurements for each star in Sect. 5. Finally, in Sect. 6 we discuss the result from this study.


\section{The SENS program}
Our survey targets are selected from the \emph{HIPPARCOS} catalogue (ESA 1997) according to the following criteria:
stars with
1) the declination $\delta$ $\geq$ 70,
2) 5.0 < $\textit{$m_{v}$}$ < 7.0 to attain a sufficient S/N ratio, which are barely observable using the 2 m class telescopes with a Doppler precision of $\sim$10 m s$^{-1}$,
3) a color index of 0.6 < \emph{B$-$V} < 1.6 to study precise RV measurements with an ample number
of spectral lines, and
4) excluding stars identified as known photometric variables or having photometric scatters over 0.05 magnitude based on the results of the \emph{HIPPARCOS} photometry.
In total, we have selected about 200 stars for the survey, which consist of late-F, G, K, and early-M stars covering all luminosity classes (the fractions in these spectral classes are 2\%, 33\%, 60\%, and 5\%, respectively).
Our target list includes two known planetary systems. This overlap will serve as a check on the systematic difference with another RV measurements.

Our survey has three advantages.
First, the area was not actively observed for exoplanet survey.
Second, since the candidates are concentrated around the pole star, this reduces the telescope travel time to the successive target (less than 30 seconds). Finally, observations are possible throughout the year.

Our basic procedure is composed of the following three steps.
In the first step, we had observed a large number of candidate stars with a S/N ratio of $\sim$150 by 2012.
During this period, observing time is scheduled such that each target should receive 4$-$5 observations per year for the first screening to identify stars showing large RV variations.
This strategy would appear to reduce the probability of detecting shorter-period exoplanets because they require more densely sampled observations. We, however, note that our sample contains just 5\% of MS star, which are expected to be slightly more frequently observed.
In the second step, for stars that turn out to show large RV variations, more than three times RV variations of the RV standard star $\tau$ Ceti, we conduct follow-up observations to confirm their periodicity. This proceeds for two years.
In a final step, we focus on the detection of planets with periodicity.
We plan to monitor the selected targets during the next two years to cover a complete orbital cycle for planetary candidates. This program takes seven years in total.

%
\section{Observations and data reduction}
Observations were carried out using the fiber-fed high-resolution Bohyunsan Observatory Echelle Spectrograph (BOES; Kim et al. 2007) attached to a 1.8 m telescope at BOAO in Korea. Generally, to reduce the line broadening due to a stellar rotation, an exposure time is limited to less than about 20 minutes in the precise RV measurements.
In order to obtain an appropriate signal with respect to the candidates (5.0 < $\textit{$m_{v}$}$ < 7.0), the resolution of the optical fiber of 45 000 is most advantageous among three kinds of resolutions R = 90 000, 45 000, and 27 000 available.
To obtain precise RV measurements, we used an iodine absorption (I$_{2}$) cell, which superimposes thousands of molecular absorption lines over the object spectra in the spectral region between 4900 and 6100 ${\AA}$. Using these lines as a wavelength standard, we simultaneously model the time-variant instrumental profile and Doppler shift relative to an I$_{2}$ free template spectrum.

Observations for the SENS began at BOAO in January 2010. The program has received about 17 nights per year from the beginning, and approximately 40\% has produced usable data. We aim for a S/N ratio of $\sim$150 at 5500 ${\AA}$, resulting in exposure times ranging from 10 to 20 minutes.

Extraction of the spectra from raw CCD images was carried out using the IRAF software that performs bias correction, flat fielding, removing scattered light from inter-order pixels, and subtracts one dimensional spectra. The precise RV measurements related to the I$_{2}$ analysis were undertaken using the RVI2CELL (Han et al. 2007), which is based on a method by Butler et al. (1996) and Valenti et al. (1995).
The long-term stability of the BOES has been monitored  by observing the RV standard star $\tau$ Ceti since 2003. It demonstrates a long-term RV stability of the spectrometer with an rms scatter of about 7 m~s$^{-1}$ (Lee et al. 2013).

%
\section{Stellar characteristics}

\subsection{Fundamental parameter}
The basic stellar parameters were taken from the \emph{HIPPARCOS} catalog (ESA 1997) and improved parallaxes from van Leeuwen (2007). The distance and luminosity came from the result of Anderson \& Francis (2012).
The stellar atmospheric parameters were determined using the TGVIT (Takeda et al. 2005) program based on the Kurucz (1992) atmosphere models.
We used more than 150 equivalent width (EW) measurements of Fe~I and Fe~II lines each star.
Stellar radii and masses were estimated from the stellar  positions in the color$-$magnitude diagram and by using the theoretical stellar isochrones of Bressan et al. (2012).
We also adopted a version of the Bayesian estimation method (J{\o}rgensen \& Lindegren 2005; da Silva et al. 2006) by using the determined values for $T_{\mathrm{eff}}$, $\mathrm{[Fe/H]}$, $m_{v}$, and $\pi$.
It also provided the most likely values of stellar age. The basic stellar parameters are summarized in Table~\ref{tab1}.

\subsection{Rotational period}
In evolved stars, the stellar rotational period is very important in identifying the RV variation from the rotational modulations of surface structures. The observed stellar spectrum was fitted by convolving the component functions, which broaden spectral lines without altering their EW. Determining the origin of the broadening is difficult for spectra of slowly rotating late type stars because they have similar intrinsic line profiles.

To estimate the stellar rotational velocities, a line-broadening model by Takeda et al. (2008) was used.
For the determination of line broadening, the automatic spectrum-fitting technique (Takeda 1995) for the spectrum within the wavelength range  6080$-$6089 {\AA} was also used.
We estimated the rotational velocities of 2.3 km~s$^{-1}$ for HD~11755, 4.8 km~s$^{-1}$ for HD~12648, 3.5 km~s$^{-1}$ for HD~24064, and 3.6 km~s$^{-1}$ for 8~UMi. Based on the rotational velocities and the stellar radii, we derived the upper limits for the rotational period of 600.5 days for HD~11755, 97.0 days for HD~12648, 549.3 days for HD~24064, and 139.1 days for 8~UMi.

   \begin{figure}
   \centering
   \includegraphics[width=7cm]{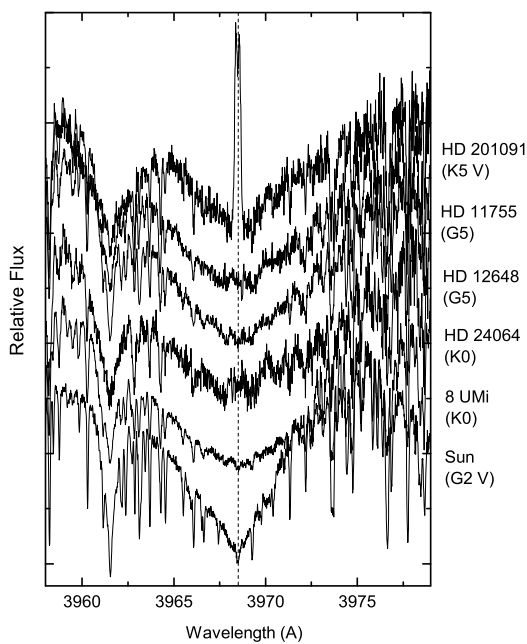}
      \caption{Ca II H line cores for our program stars including the chromospheric active star HD~201091 and the Sun. There are no core reversals in the center of the Ca II H line for three stars except HD~24064. The vertical dotted line indicates the center of the Ca II H regions.
        }
        \label{Ca1}
   \end{figure}
%

%
\begin{table*}
\begin{center}
\caption[]{Stellar parameters for the stars analyzed in the present paper.}
\label{tab1}
\begin{tabular}{lccccc}
\hline
\hline
    Parameters                & HD 11755          & HD 12648          & HD 24064          & 8 UMi              & Ref.     \\

\hline
    $\alpha$ (J2000)                & 01 58 50.087      & 02 18 59.654      & 03 56 36.297      & 14 56 48.353       & 1 \\
    $\delta$ (J2000)                & +73 09 08.58      & +85 44 10.22      & +74 04 48.12      & +74 54 03.34       & 1 \\
    Spectral type             & G5                & G5                & K0                & K0                 & 1 \\
    $\textit{$m_{v}$}$ (mag)  & 6.87              & 6.98              & 6.75              & 6.83               & 1 \\
    \emph{HIPPARCOS}$_{\rm scat}$     (mag)      & 0.010             & 0.009             & 0.009             &  0.008             & 1 \\
    $\emph{B$-$V}$      (mag)  & 1.216 $\pm$ 0.008 & 0.904 $\pm$ 0.008 & 1.513 $\pm$ 0.008 & 0.985 $\pm$ 0.010  & 1 \\
    Parallax           (mas)  & 4.27 $\pm$ 0.42    & 6.29$\pm$ 0.36    & 3.72 $\pm$ 0.44    & 6.25 $\pm$ 0.43    &  2 \\
    Distance           (pc)   & 231.5 $\pm$ 22.7   & 158.4 $\pm$ 9.0   & 264.4 $\pm$ 31.2   & 159.1 $\pm$ 11.0   & 3 \\
    Age                (Gyr)  & 10.2 $\pm$ 1.3     & 4.5 $\pm$ 1.0     & 9.0 $\pm$ 2.1     & 1.7 $\pm$ 0.2     &   4\tablefootmark{a} \\
    $T_{\mathrm{eff}}$  (K)   & 4312.5 $\pm$ 5.0   & 4835.8 $\pm$ 7.5   & 4052.5 $\pm$ 22.5  & 4847.4 $\pm$ 7.5    & 4 \\
    $\mathrm{[Fe/H]}$         & $-$0.74 $\pm$ 0.02 & $-$0.57 $\pm$ 0.02 & $-$0.49 $\pm$ 0.06 & $-$0.03 $\pm$ 0.02  & 4 \\
    log $\it g$               & 1.67 $\pm$ 0.03    & 2.18 $\pm$ 0.03    & 1.44 $\pm$ 0.11    & 2.57 $\pm$ 0.03     & 4 \\
    $\textit{$R_{\star}$}$ ($R_{\odot}$) & 27.3 $\pm$ 1.0 & 9.2 $\pm$ 0.6 & 38.0 $\pm$ 2.9 & 9.9 $\pm$ 0.4  & 4\tablefootmark{a}  \\
    $\textit{$M_{\star}$}$ ($M_{\odot}$) & 0.9 $\pm$ 0.1  & 1.2 $\pm$ 0.1 & 1.0 $\pm$ 0.1  & 1.8 $\pm$ 0.1  & 4\tablefootmark{a}  \\
    $\textit{$L_{\star}$}$ [$L_{\odot}$] & 145.72 & 45.01 & 352.90 &  55.94 &  3 \\
    $v_{\mathrm{rot}}$ sin $i$ (km s$^{-1}$) & 2.3             & 4.8             & 3.5             & 3.6             & 4\tablefootmark{b}   \\
    $P_{\mathrm{rot}}$ / sin $i$ (days)      & 600.5           & 97.0            & 549.3           & 139.1           & 4\tablefootmark{b}   \\
    $v_{\mathrm{micro}}$ (km s$^{-1}$)       & 1.50 $\pm$ 0.04 & 1.52 $\pm$ 0.03 & 1.53 $\pm$ 0.09 & 1.48 $\pm$ 0.04 & 4                    \\

\hline

\end{tabular}
\tablebib{(1) \emph{HIPPARCOS}; (2) van Leeuwen (2007); (3) Anderson \& Francis (2012); (4) This work.}
\tablefoot{
\tablefoottext{a}{Derived using the online tool (http://stevoapd.inaf.it/cgi-bin/param).}
\tablefoottext{b}{See text.}
}
\end{center}
\end{table*}
\subsection{Photometric variations}
The \emph{HIPPARCOS} photometry data obtained from December 1989 to February 1993 (HD~11755), from December 1989 to March 1993 (HD~12648), from December 1989 to February 1993 (HD~24064), and from February 1990 to March 1993 (8~UMi) were analyzed to find possible brightness variations that may be caused by the rotational modulation of large stellar spots. This database yields 149, 143, 162, and 111 measurements for HD 11755, HD~12648, HD~24064, and 8~UMi, respectively. All stars were photometrically constant down to an rms scatter of 0.008$-$0.010 magnitude. This  corresponds to a variation over the time span of the observations of only 0.14\%, 0.13\%, 0.13\% and 0.12\% for the four stars, respectively.

\subsection{Chromspheric activity}
Since the first discovery of emission lines in the core of the Ca~II absorption feature (Eberhard \& Schwarzschild 1913), the EW variations of the Ca II H have been frequently used as a chromospheric activity indicator.
The Ca II emission feature at the line center implies that the source function in the chromosphere is greater than that in the photosphere.
This is common in cool stars and is directly connected to the convective envelope and magnetic activity.
Stellar activity can derive RV variations that can mask or even mimic the RV signature of orbiting planetary companions.
Figure~\ref{Ca1} shows the Ca II H line region of the six BOES spectra and the chromospheric active star HD~201901 (K5 V) and the Sun (G2 V) are shown for comparisons.
The star 8~UMi lacks a prominent core emission feature in Ca II H and two stars HD~11755 and HD~12648 are chromospherically modest.
Unfortunately, the spectrum is not clear enough to resolve the emission feature for HD~24064. There may be a slight central emission in HD~24064 indicating a low level of stellar activity.

To clear up the ambiguity over chromospheric activity, we also measured variaitons of H$_{\alpha}$ EW for the sample.
The EW variations of the H$_{\alpha}$ line are frequently used as the chromospheric activity indicator along with Ca II H line (Montes et al. 1995; K{\"u}rster et al. 2003; Lee et al. 2012; Hatzes et al. 2015).
We measured the EW using a band pass of $\pm$ 1.0 {\AA} centered on the core of the H$_{\alpha}$ line to avoid nearby blending lines (i.e. ATM H$_{2}$O 6561.1, Ti I 6561.3, Na II 6563.9, and ATM H$_{2}$O 6564.2 {\AA}).
The individual chromospheric activity measurements of the H$_{\alpha}$ line in Lomb-Scargle (L-S) periodograms are shown in the third panel of Figs.~\ref{power1}, \ref{power2}, \ref{power3}, and \ref{power4}.

\subsection{Bisector variations}
Stellar rotational modulations by inhomogeneous surface features can create variable asymmetries in the spectral line profiles (Queloz et al. 2001). The differential RV measurements between the high and low flux levels of the line profile are defined as a bisector velocity span (BVS = V$_{top}$ $-$ V$_{bottom}$). The changes in the spectral line bisector can also be quantified using the bisector velocity curvature (BVC = [V$_{top}$ $-$ V$_{center}$] $-$ [V$_{center}$ $-$ V$_{bottom}$]).
A BVS is simply the velocity difference in the bisectors of line widths between the top and bottom of the line profile and a BVC is the difference of the velocity span of the upper half and the lower half of the bisector.

For searching variations in the spectral line shapes, we select Ni I 6643.6 {\AA} line as described in Hatzes et al. (2005) and Lee et al. (2013; 2014a), which is an unblended spectral feature with a high flux level and is located beyond the I$_{2}$ absorption region, so contamination should not affect our bisector measurements. We measured the bisector variations of the profile between two different flux levels at 0.8 and 0.4 of the central depth as the span points, thereby avoiding the spectral core and wing where errors of the bisector measurements are large.
The individual measurements for line bisectors are shown in the bottom panel of Figs.~\ref{power1}, \ref{power2}, \ref{power3}, and \ref{power4} (see the text).

%
\section{Radial velocity variations and their origins}
In order to search for periodicity in observed RV time series data, we performed the L-S periodogram analysis (Scargle 1982).
This is a useful tool to investigate long-period variations for unequally spaced data.
The resulting RV measurements are listed as online data (Tables \ref{tab3}$-$\ref{tab6}).

\subsection{HD 11755}
%

   \begin{figure}
   \centering
   \includegraphics[width=8cm]{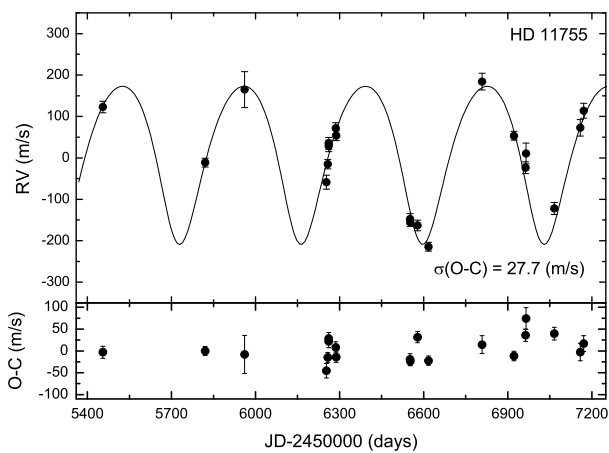}
      \caption{The RV measurements for HD 11755 from September 2010 to May 2015. (\emph{top panel}) Observed RVs for HD 11755 and the solid line is a Keplerian orbital fit with a period of 433.7 days, a semi-amplitude of 191.3 m s$^{-1}$, and an eccentricity of 0.19, yielding a minimum companion mass of 6.5 $M_{\rm{Jup}}$. (\emph{bottom panel}) Residual velocities remaining after subtracting the companion model from observations.
              }
         \label{rv1}
   \end{figure}
%

%
 \begin{figure}
   \centering
   \includegraphics[width=8cm]{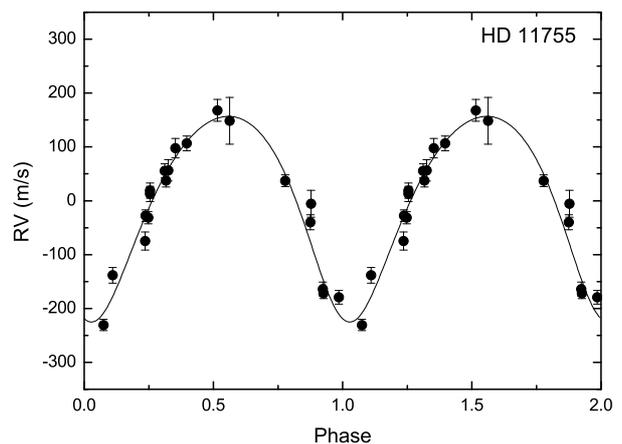}
      \caption{The RV variations for HD 11755 phased with the orbital period of 433.7 days.
        }

        \label{phase1}
   \end{figure}
%

   \begin{figure}
   \centering
   \includegraphics[width=8cm]{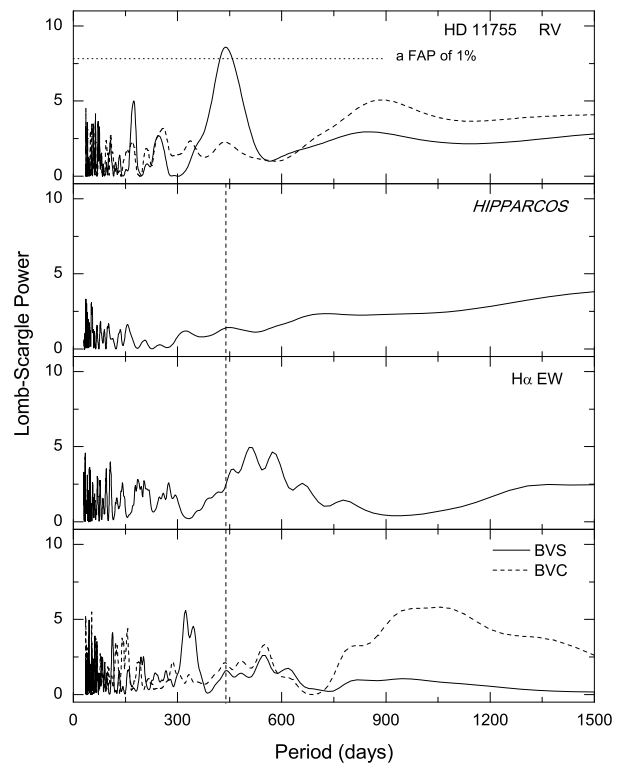}
      \caption{The L-S periodograms of the RV measurements, the \emph{HIPPARCOS} photometric measurements, the H$_{\alpha}$ EW variations, and the bisector variations for HD 11755 (\emph{top} to \emph{bottom panel}). The vertical dashed line marks the location of the period of 433 days.
      (\emph{top panel}) The solid line is the L-S periodogram of the RV measurements for five years and the periodogram shows a significant power at a period of 433.7 days. The dashed line is the periodogram of the residuals after removing of the main period fit from the original data. The horizontal dotted line indicates a FAP threshold of 1 $\times 10^{-2}$ (1\%).
        }
        \label{power1}
   \end{figure}

Twenty BOES spectra observations for HD 11755 spanning five years have been obtained, as shown in Fig.~\ref{rv1} and Table~\ref{tab3}. The observations span four full orbital periods. The L-S periodogram of the RV measurements for HD 11755 shows a significant peak at a period of 433.7 days (\emph{top panel} in Fig.~\ref{power1}). The L-S power of this peak corresponds to a false alarm probability (FAP) of $<$ 1 $\times$ 10$^{-3}$, adopting the procedure described in Cumming (2004).
We found that a semi-amplitude $K$ = 191.3 $\pm$ 10.2 m s$^{-1}$ and an eccentricity $e$ = 0.19 $\pm$ 0.10 for a Keplerian orbital fit. The rms of the RV residuals to the Keplerian orbital fit is 27.7 m s$^{-1}$.  We calculated secondary period in the residual for HD 11755 and it shows no significant peak (dashed line of \emph{top panel} in Fig.~\ref{power1}).
Adopting a mass of 0.9 $\pm$ 0.1 $M_{\odot}$ for HD 11577, we fit the companion mass \emph{m}~sin~$i$ = 6.5 $\pm$ 1.0 $M_{\rm{Jup}}$ at a distance of 1.08 $\pm$ 0.04 AU from the host star.

In order to check for the stellar brightness variations on HD~11755, the \emph{HIPPARCOS} photometric data were analyzed, which shows no correlations with RV variations (\emph{second panel} in Fig.~\ref{power1}).
To examine for any RV fluctuations that would be induced by stellar activity, the H$_{\alpha}$ EW variations were examined. Figure~\ref{power1} shows the L-S periodgram of the H$_{\alpha}$ EW variations (\emph{third panel}). No correlation was found between the H$_{\alpha}$ EW and the RV variations.
This means that HD 11755, at the moment of observations, exhibited at most a modest chromospheric activity.
Finally, line bisectors were measured by two kind of methods. No correlation was found between the RV and the BVS variations nor between the RV and the BVC variations (\emph{bottom panel} in Fig.~\ref{power1}). All Keplerian orbital elements are listed in Table~\ref{tab2}.

\subsection{HD 12648}
%

   \begin{figure}
   \centering
   \includegraphics[width=8cm]{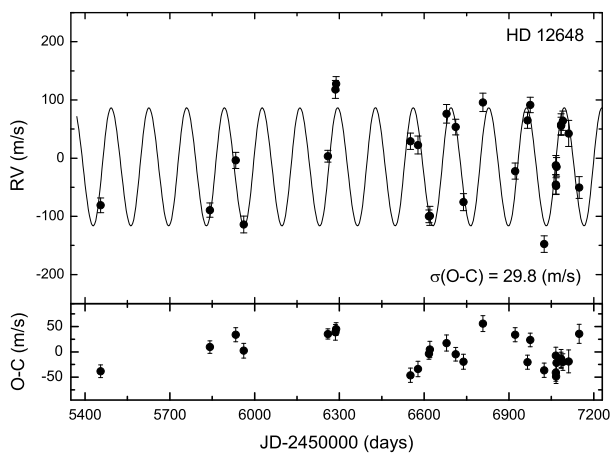}
      \caption{The RV measurements for HD 12648 from September 2010 to May 2015. (\emph{top panel}) Observed RVs for HD 12648 and the solid line is a Keplerian orbital fit with a period of 133.6 days, a semi-amplitude of 102.0 m s$^{-1}$, and an eccentricity of 0.04, yielding a minimum companion mass of 2.9 $M_{\rm{Jup}}$. (\emph{bottom panel}) Residual velocities remaining after subtracting the companion model from observations.
              }
         \label{rv2}
   \end{figure}
 \begin{figure}
   \centering
   \includegraphics[width=8cm]{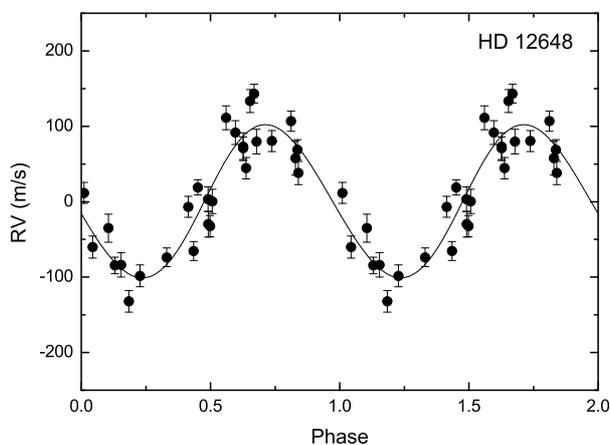}
      \caption{The RV variations for HD 12648 phased with the orbital period of 133.6 days.
        }

        \label{phase2}
   \end{figure}
%

   \begin{figure}
   \centering
   \includegraphics[width=8cm]{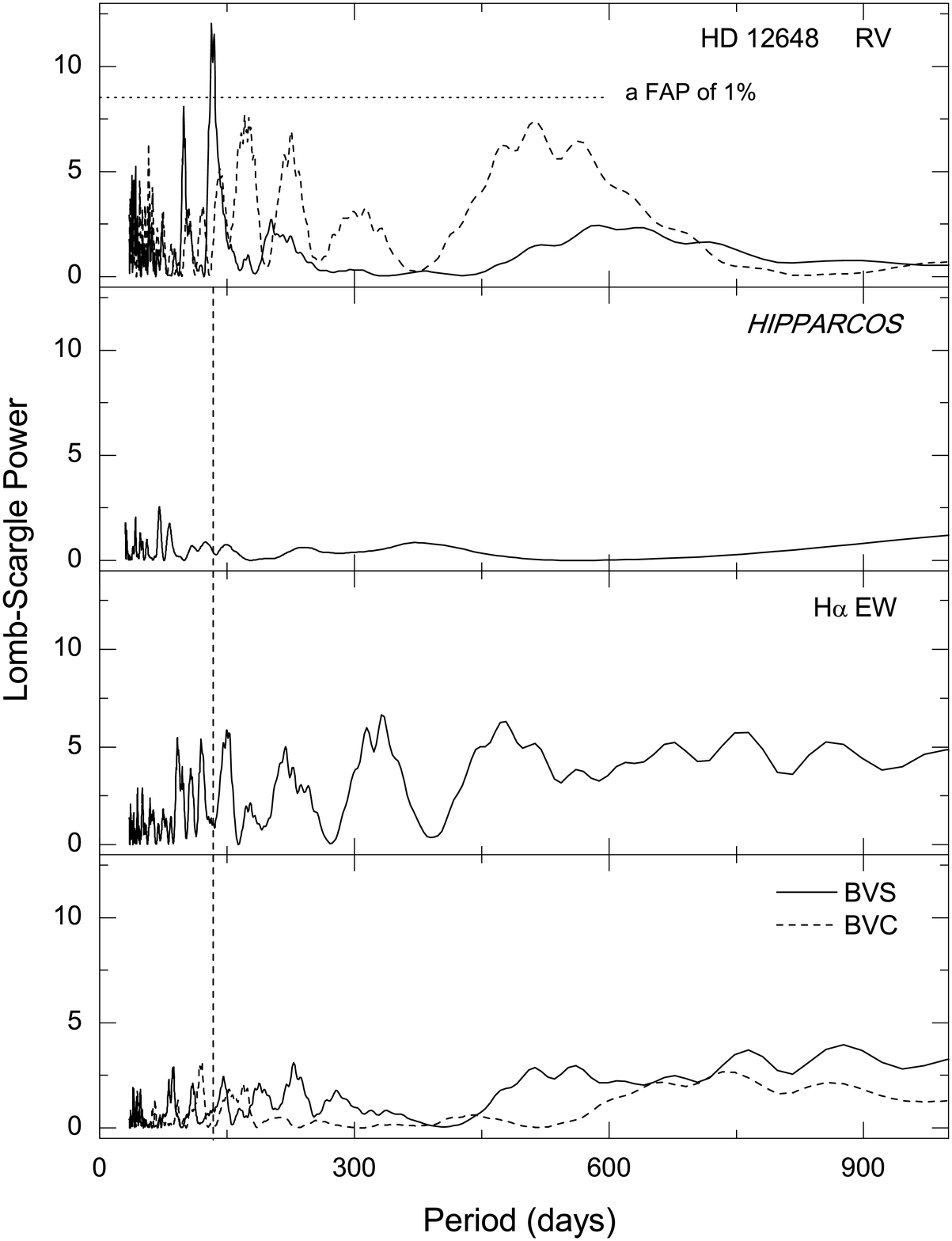}
      \caption{The L-S periodograms of the RV measurements, the \emph{HIPPARCOS} photometric measurements, the H$_{\alpha}$ EW variations, and the bisector variations for HD 12648 (\emph{top} to \emph{bottom panel}). The vertical dashed line marks the location of the period of 133 days.
      (\emph{top panel}) The solid line is the L-S periodogram of the RV measurements for five years and the periodogram shows a significant power at a period of 133.6 days. The dashed line is the periodogram of the residuals after removing of the main period fit from the original data. The horizontal dotted line indicates a FAP threshold of 1 $\times 10^{-2}$ (1\%).
        }
        \label{power2}
   \end{figure}

Observations for HD 12648 with the BOES took place between September 2010 and May 2015. We have gathered a total of 28 data points during five years (Fig.~\ref{rv2} and Table~\ref{tab4}). We found that the primary RV variations were fitted best with a Keplerian orbit with a period $P$ = 133.6 $\pm$ 0.5 days, a semi-amplitude $K$ = 102.0 $\pm$ 8.4 m s$^{-1}$, and an eccentricity $e$ = 0.04 $\pm$ 0.16. The rms of the RV residuals to the Keplerian orbital fit is 29.8 m s$^{-1}$.  We calculated a secondary period in the residual for HD 12648 and it shows no significant peak (dashed line of \emph{top panel} in Fig.~\ref{power2}).
Assuming a mass $M_{\star}$ = 1.2 $\pm$ 0.1 $M_{\odot}$ for HD 12648, the result implies that the companion mass is \emph{m}~sin~$i$ = 2.9 $\pm$ 0.4 $M_{\rm{Jup}}$ at a distance of 0.54 $\pm$ 0.02 AU from the host star.

The L-S periodograms of the \emph{HIPPARCOS} measurements, the H$_{\alpha}$ EW variations, and line bisector variations for HD 12648 are shown in Fig.~\ref{power2}. They do not show any obvious correlations with the RV measurements. This suggests that RV variations are not caused by line-shape changes produced by rotational modulation of surface features but by the planetary companion.

\subsection{HD 24064}
%

   \begin{figure}
   \centering
   \includegraphics[width=8cm]{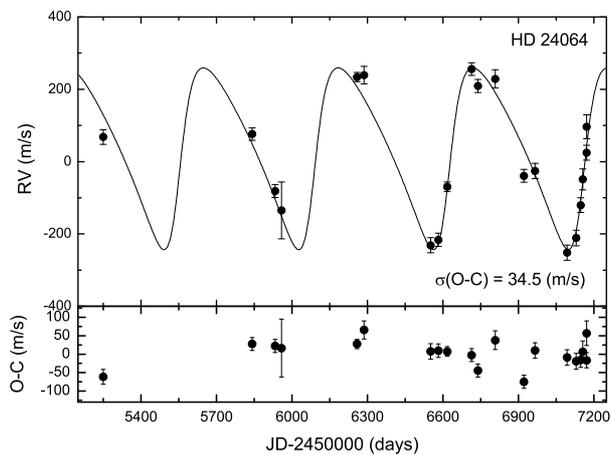}
      \caption{The RV measurements for HD 24064 from February 2010 to May 2015. (\emph{top panel}) Observed RVs for HD 24064 and the solid line is a Keplerian orbital fit with a period of 535.6 days, a semi-amplitude of 251.0 m s$^{-1}$, and an eccentricity of 0.35, yielding a minimum companion mass of 9.4 $M_{\rm{Jup}}$. (\emph{bottom panel}) Residual velocities remaining after subtracting the companion model from observations.
              }
         \label{rv3}
   \end{figure}
%

%
 \begin{figure}
   \centering
   \includegraphics[width=8cm]{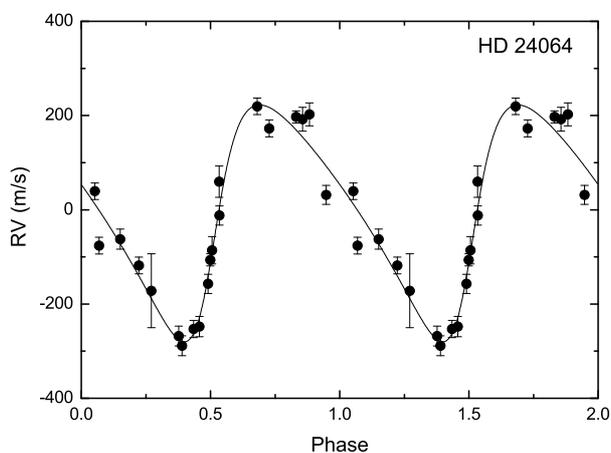}
      \caption{The RV variations for HD 24064 phased with the orbital period of 535.6 days.
        }

        \label{phase3}
   \end{figure}
%
   \begin{figure}
   \centering
   \includegraphics[width=8cm]{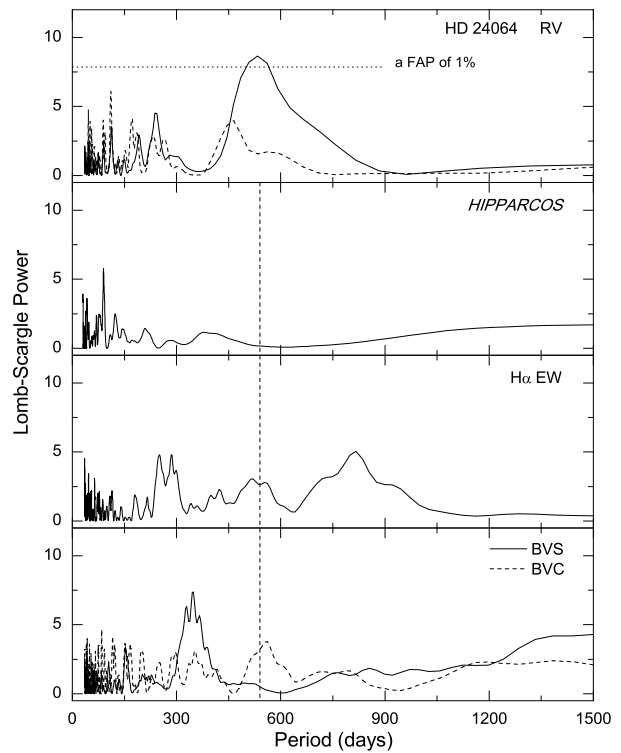}
      \caption{The L-S periodograms of the RV measurements, the \emph{HIPPARCOS} photometric measurements, the H$_{\alpha}$ EW variations, and the bisector variations for HD 24064 (\emph{top} to \emph{bottom panel}). The vertical dashed line marks the location of the period of 535 days.
      (\emph{top panel}) The solid line is the L-S periodogram of the RV measurements for five years and the periodogram shows a significant power at a period of 535.6 days. The dashed line is the periodogram of the residuals after removing of the main period fit from the original data.      The horizontal dotted line indicates a FAP threshold of 1 $\times 10^{-2}$ (1\%).
        }
        \label{power3}
   \end{figure}

Since February 2010, we have gathered 20 spectra for HD~24064 displaying in Fig.~\ref{rv3} and listed at Table~\ref{tab5}. The observations span three and half orbital periods. The L-S periodogram of the RV measurements for HD 24064 shows a significant peak at a period of 535.6 days (\emph{top panel} in Fig.~\ref{power3}). The L-S power of this peak corresponds to a FAP of $<$ 1 $\times$ 10$^{-3}$. We found that a semi-amplitude $K$ = 250.8 $\pm$ 6.3 m~s$^{-1}$ and an eccentricity $e$ = 0.35 $\pm$ 0.08. After removing the main signal, the dispersion of the RV residuals is 34.5 m~s$^{-1}$, which is significantly higher than the RV precision for the RV standard star $\tau$~Ceti ($\sim$7~m~s$^{-1}$) or the typical internal error of individual RV accuracy of $\sim$13.8~m~s$^{-1}$ for HD 24064. A periodogram of the RV residual, however, does not show any additional periodic signal as shown in Fig.~\ref{power3} (\emph{top panel}).
Assuming a mass $M_{\star}$ = 1.0 $\pm$ 0.1 $M_{\odot}$, we fit a companion mass \emph{m}~sin~$i$ = 9.4 $\pm$ 1.3 $M_{\rm{Jup}}$ at a distance of 1.29 $\pm$ 0.05 AU from HD~24064.

The L-S periodograms of the \emph{HIPPARCOS} measurements, the H$_{\alpha}$ EW variations, and line bisector variations for HD~24064 are shown in Fig.~\ref{power3}. The H$_{\alpha}$ EW does not show any obvious correlations with the RV measurements. We find a FAP of 18\% for the peak at 89.8 days in the \emph{HIPPARCOS} measurements and the dominant peak of BVS is near 350 days with a FAP over 3\%, but they show no correlation to the RV. Even though there is a peak in the L-S periodogram of BVC roughly at 557 days, the FAP is $\sim$ 27\% and, therefore, it is not statistically meaningful. While this star shows a bit of fluctuations in chromospheric activity, it is evident from Fig.~\ref{power3} that stellar activities are uncorrelated with the RV measurements. Furthermore, the periodograms of the bisector also indicates no periodicity near the 535-day period of the planetary companion. These independent lines of evidence thus lead us to conclude that the RV variations observed in HD 24064 are not attributable to an intrinsic stellar process but to an orbiting giant planet.

\subsection{8 Ursae Minoris}
%
   \begin{figure}
   \centering
   \includegraphics[width=8cm]{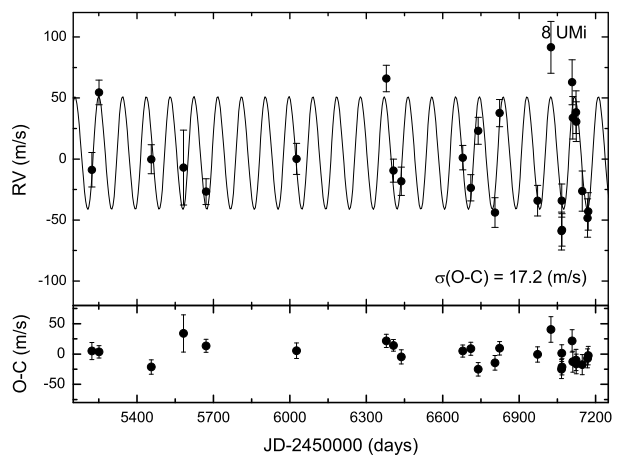}
      \caption{The RV measurements for 8 UMi from January 2010 to May 2015. (\emph{top panel}) Observed RVs for 8 UMi and the solid line is a Keplerian orbital fit with a period of 93.4 days, a semi-amplitude of 46.1 m s$^{-1}$, and an eccentricity of 0.06, yielding a minimum companion mass of 1.5 $M_{\rm{Jup}}$. (\emph{bottom panel}) Residual velocities remaining after subtracting the companion model from observations.
              }
         \label{rv4}
   \end{figure}
%

%
 \begin{figure}
   \centering
   \includegraphics[width=8cm]{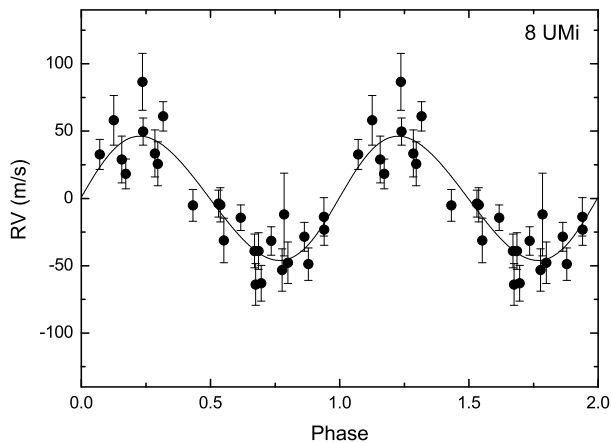}
      \caption{The RV variations for 8UMi phased with the orbital period of 93.4 days.
        }

        \label{phase4}
   \end{figure}
%

   \begin{figure}
   \centering
   \includegraphics[width=8cm]{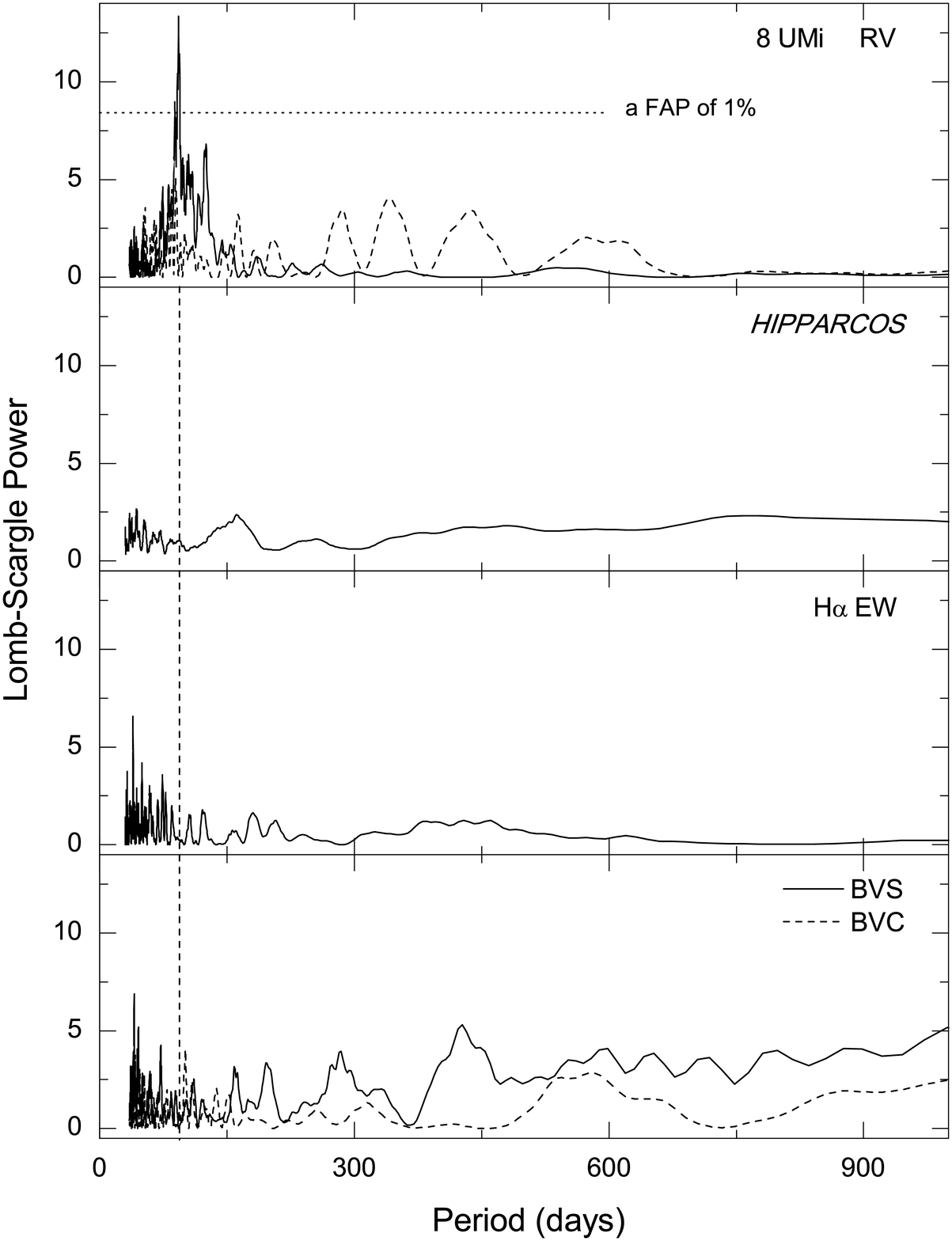}
      \caption{L-S periodograms of the RV measurements, the \emph{HIPPARCOS} photometric measurements, the H$_{\alpha}$ EW variations, and the bisector variations for 8 UMi (\emph{top} to \emph{bottom panel}). The vertical dashed line marks the location of the period of 93 days.
      (\emph{top panel}) The solid line is the L-S periodogram of the RV measurements for five years and the periodogram shows a significant power at a period of 93.4 days. The dashed line is the periodogram of the residuals after removing of the main period fit from the original data. The horizontal dotted line indicates a FAP threshold of 1 $\times 10^{-2}$ (1\%).
        }
        \label{power4}
   \end{figure}
%

\begin{table*}
\begin{center}
\caption{Orbital parameters for four exoplanets.}
\label{tab2}
\begin{tabular}{lcccc}
\hline
\hline
    Parameter                        & HD 11755 b         & HD 12648 b         & HD 24064 b          & 8 UMi b            \\
\hline
    P (days)                         & 433.7 $\pm$ 3.2    & 133.6 $\pm$ 0.5    & 535.6 $\pm$ 3.0     & 93.4  $\pm$ 4.5    \\
    $\it T$$_{\rm{periastron}}$ (JD) & 2457018.2$\pm$14.0 & 2452324.7$\pm$57.6 & 2455278.3$\pm$11.8  & 2454108.5$\pm$22.8 \\
    $\it{K}$ (m s$^{-1}$)            & 191.3 $\pm$ 10.2   & 102.0  $\pm$ 8.4   & 251.0  $\pm$ 9.3    & 46.1 $\pm$ 4.0   \\
    $\it{e}$                         & 0.19 $\pm$ 0.10    & 0.04 $\pm$ 0.16    & 0.35 $\pm$ 0.08     & 0.06 $\pm$ 0.18   \\
    $\omega$ (deg)                   & 155.3 $\pm$ 12.6   & 99.3 $\pm$ 140.7   & 250.8 $\pm$ 6.3     & 91.0 $\pm$ 84.6  \\
    \emph{m}~sin~$\it i$ ($M_{Jup}$) & 6.5 $\pm$ 1.0      & 2.9 $\pm$ 0.4      & 9.4 $\pm$ 1.3       & 1.5 $\pm$ 0.2    \\
    $\it{a}$ (AU)                    & 1.08 $\pm$ 0.04    & 0.54  $\pm$ 0.02   & 1.29  $\pm$ 0.05    & 0.49  $\pm$ 0.03   \\
    $N_{obs}$                        & 20                 &  28                & 20                  &  26              \\
    rms (m s$^{-1}$)                 & 27.7               &  29.8              & 34.5                &  17.2            \\
\hline

\end{tabular}
\end{center}
\end{table*}

We obtained 26 BOES spectra for 8 UMi spanning five years (Fig.~\ref{rv4} and Table~\ref{tab6}). The L-S periodogram of the RV measurements for 8 UMi shows a significant peak at a period of 93.4 $\pm$ 4.5 days (\emph{top panel} in Fig.~\ref{power4}). We found that a semi-amplitude $K$ = 46.1 $\pm$ 4.0 m s$^{-1}$ and an eccentricity $e$ = 0.06 $\pm$ 0.18 for the Keplerian orbital fit. The rms of the RV residuals to the Keplerian orbital fit is 17.2 m s$^{-1}$.
By adopting a stellar mass of 1.8 $\pm$ 0.1 $M_{\odot}$ for 8 UMi, we obtain a minimum companion mass of 1.5 $\pm$ 0.2 $M_{\rm{Jup}}$ and a semi-major axis of 0.49 $\pm$ 0.03 AU.

The \emph{HIPPARCOS} measurements, the H$_{\alpha}$ EW variations, and line bisector variations for 8 UMi show no correlations with the RV measurements. The lack of any significant peaks in the L-S periodograms of chromospheric activity indicators shows that RV measurements of 8 UMi is caused by a planetary companion.

\section{Discussion}
Using the BOAO/BOES, we have begun a search for exoplanets around circumpolar stars in the northern hemisphere.
All of our stars have been surveyed for six years and $\sim$16\% of them revealed periodic RV variations. This is consistent with a prediction that about 18\% of F, G, and K stars harbor a planet or candidate (Cumming et al. 2008).

We found periodic RV variations in HD 11755, HD 12648, HD 24064, and 8 UMi.
They are evolved stars that are currently in the giant stage with stellar classifications of G0 and K0.
Since most of giants have intrinsic RV variations of several hundred days (Hekker et al. 2008), it is necessary to critically examine the planet hypothesis for these stars. Generally, most of them exhibit pulsations and/or surface activities with RV variabilities on different time scales. To establish the origin of the pure RV variabilities, some relevant analyses were carried out: rotational periods, photometric data, Ca II H line inspection, H$_{\alpha}$ EW measurements, and line bisector measurements.

The simplest test is to obtain the maximum rotational period.
Based on the rotational velocities and the stellar radii, we derived the upper limits for the rotational period of 600.5 days for HD~11755, 97.0 days for HD~12648, 549.3 days for HD~24064, and 139.1 days for 8~UMi. Of these, because the RV period of HD~12648 (133 days) is larger than the upper limit of the rotational period of 97 days, observed RV variations in HD~12648 cannot be associated with the stellar rotation.
Although the other three show no strong correlations between the RV variations and rotational periods, we cannot exclude the rotational modulations. We just note that the detected RV periods can be compatible with the rotational periods of the sample, even though that is not the most likely explanation.

However, the examination of the activity indicator Ca II H line does not show any obvious evidence of chromospheric activity for our sample. While a weak central emission may exist in the center of Ca II H line for HD~24064, which is not strong enough to warrant the presence of the chromospheric active. From another measurements of chromospheric activity (H$_{\alpha}$ EW variations), no significant peaks are evident, even in HD~24064.
We can thus exclude that the main periodic RV variations observed in HD~24064 are induced by stellar activity.
We identified the \emph{HIPPARCOS} photometries and calculated periodograms. None of the stars show photometric variations related to the observed RV variations. Line bisector analysis is a fairly common technique to determine if periodic RV variations are caused by the rotational modulation of stellar spots. Such analysis has the advantage of being contemporaneous with the RV measurements. No traces of line bisector variations were found associated with the RV variations.

\begin{figure}
   \centering
   \includegraphics[width=8cm]{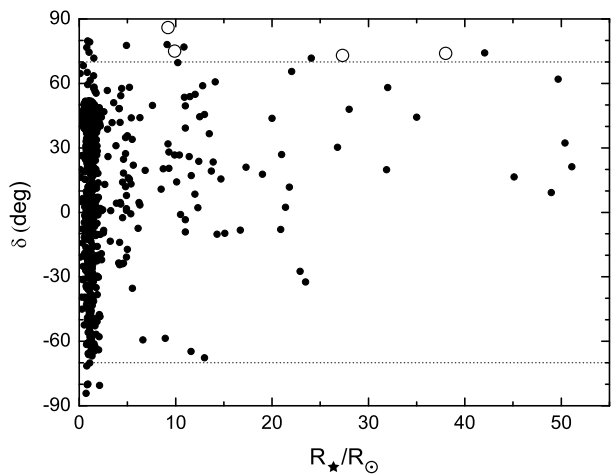}
      \caption{As of July 2015, distribution of planetary companions between the stellar radii and declinations. Closed circles are known exoplanets and open circles denote the locations of new four exoplanets HD~12648, 8~UMi, HD~11755, and HD~24064 (left to right). The horizontal dotted lines indicate a declination of $\pm$ 70\,$^{\circ}$.
        }

        \label{stat}
   \end{figure}

We thus conclude that giants HD~11755, HD~12648, HD~24064, and 8~UMi host a planetary companion with a period of 433, 133, 535, and 93 days and a minimum mass of 6.5, 2.9, 9.4, and 1.5 $M_{Jup}$, respectively.
Figure~\ref{stat} shows the distribution of the stellar radii versus the declinations of the exoplanets discovered so far.
HD~12648~b becomes the exoplanet detected in the highest declination of 85.7\,$^{\circ}$ so far and HD~24064~b corresponds to an exoplanet found around one of the largest stars.
The planets discovered in five of the six stars larger than HD~24064 were reported in another exoplanet survey program using 1.8~m telescope at BOAO (Lee et al. 2013; Lee et al. 2014b; Hatzes et al. 2015).

In summary, we found four new exoplanets around giant stars by expanding the exoplanet survey region to be the north pole region (the declination $\delta$ $\geq$ 70). About 20 exoplanets (corresponding to $\sim$1\% of the total) have been found around the pole region so far. Interestingly, nearly half have been found around giant stars. These discoveries will be important in understanding the planet formation around giant stars and the effect of the stellar evolution on the planet formation. Additional observations including south pole region can confirm the statistical characteristics of these planetary systems.

%
%


\begin{acknowledgements}
      BCL, JWL, CUL, and SLK acknowledge partial support by the KASI (Korea Astronomy and Space Science Institute) grant 2015-1-850-04. Support for MGP was provided by the National Research Foundation of Korea to the Center for Galaxy Evolution Research (No. 2012-0027910). This research made use of the SIMBAD database, operated at the CDS, Strasbourg, France. We thank the developers of the Bohyunsan Observatory Echelle Spectrograph (BOES) and all staff of the Bohyunsan Optical Astronomy Observatory (BOAO).
\end{acknowledgements}
%


\Online

\begin{table}
\begin{center}
\caption{RV measurements for HD 11755 from September 2010 to May 2015 using the BOES.}
\label{tab3}
\begin{tabular}{crr|crr}
\hline\hline JD         & $\Delta$RV  & $\pm \sigma$ &        JD & $\Delta$RV  & $\pm \sigma$  \\
$-$2450000 & m\,s$^{-1}$ &  m\,s$^{-1}$ & $-$2450000  & m\,s$^{-1}$ &  m\,s$^{-1}$  \\
\hline

5455.3146  &    122.8   &  13.7 &    6552.2523 &  $-$155.6  &    9.9  \\
5820.3073  &  $-$11.4   &  10.5 &    6578.2070 &  $-$163.0  &   12.7  \\
5961.0203  &    164.8   &  43.5 &    6616.9513 &  $-$214.5  &   10.5  \\
6253.2048  &  $-$58.5   &  16.8 &    6808.2184 &     184.0  &   20.3  \\
6258.1507  &  $-$15.2   &  11.3 &    6922.0666 &      53.4  &   10.8  \\
6261.1406  &     28.9   &  14.1 &    6964.0179 &   $-$23.7  &   14.0  \\
6261.1538  &     35.2   &  14.1 &    6965.2009 &      10.6  &   24.9  \\
6286.0018  &     71.8   &  13.1 &    7065.9562 &  $-$122.2  &   14.6  \\
6288.1130  &     53.7   &  11.7 &    7159.2983 &      72.7  &   19.8  \\
6551.1367  & $-$147.7   &  13.0 &    7171.2660 &     113.8  &   18.1  \\

\hline
\end{tabular}
\end{center}
\end{table}
%
%

\begin{table}
\begin{center}
\caption{RV measurements for HD 12648 from September 2010 to May 2015 using the BOES.}
\label{tab4}
\begin{tabular}{crr|crr}
\hline\hline JD         & $\Delta$RV  & $\pm \sigma$ &        JD & $\Delta$RV  & $\pm \sigma$  \\
$-$2450000 & m\,s$^{-1}$ &  m\,s$^{-1}$ & $-$2450000  & m\,s$^{-1}$ &  m\,s$^{-1}$  \\
\hline

5455.3263 &   $-$81.0   &    12.6  &  6808.0381  &     95.8  &   15.8  \\
5842.1888 &   $-$89.3   &    12.4  &  6922.0962  &  $-$22.3  &   13.9  \\
5933.0289 &    $-$3.8   &    13.8  &  6965.2304  &     65.0  &   13.8  \\
5961.9263 &  $-$113.9   &    14.5  &  6975.1407  &     91.3  &   13.3  \\
6259.0790 &       3.4   &    10.1  &  7024.8943  & $-$147.6  &   14.2  \\
6286.0988 &     118.0   &    15.3  &  7065.9045  &  $-$12.1  &   16.4  \\
6288.1424 &     127.8   &    12.4  &  7065.9214  &  $-$45.3  &   17.2  \\
6551.1848 &      29.0   &    14.1  &  7066.9422  &  $-$47.8  &   14.0  \\
6578.2328 &      22.6   &    15.3  &  7068.0982  &  $-$15.1  &   16.0  \\
6616.9759 &  $-$100.0   &    10.9  &  7084.0117  &     57.5  &   18.5  \\
6620.1065 &   $-$99.2   &    16.3  &  7084.0261  &     55.5  &   15.0  \\
6679.2195 &      76.2   &    15.8  &  7090.9466  &     64.3  &   16.4  \\
6711.3220 &      53.5   &    13.4  &  7111.0923  &     42.3  &   22.4  \\
6738.9917 &   $-$75.6   &    14.6  &  7147.9716  &  $-$50.6  &   18.7  \\

\hline
\end{tabular}
\end{center}
\end{table}
%
%

\begin{table}
\begin{center}
\caption{RV measurements for HD 24064 from February 2010 to May 2015 using the BOES.}
\label{tab5}
\begin{tabular}{crr|crr}
\hline\hline JD         & $\Delta$RV  & $\pm \sigma$ &        JD & $\Delta$RV  & $\pm \sigma$  \\
$-$2450000 & m\,s$^{-1}$ &  m\,s$^{-1}$ & $-$2450000  & m\,s$^{-1}$ &  m\,s$^{-1}$  \\
\hline

5250.1474  &     68.2  &   20.2 &    6739.0180  &    209.4  &   18.0  \\
5842.2870  &     76.1  &   17.4 &    6808.2670  &    228.8  &   25.2  \\
5933.2105  &  $-$81.4  &   17.8 &    6922.2232  &  $-$39.5  &   17.7  \\
5959.0512  & $-$135.1  &   78.5 &    6965.9198  &  $-$25.7  &   21.3  \\
6259.1815  &    233.8  &   12.7 &    7094.0278  & $-$251.9  &   21.0  \\
6287.0726  &    239.2  &   24.5 &    7129.9748  & $-$211.2  &   21.5  \\
6551.2761  & $-$231.4  &   21.0 &    7147.9867  & $-$120.5  &   20.3  \\
6582.2361  & $-$216.5  &   18.3 &    7155.9862  &  $-$49.2  &   28.9  \\
6617.0197  &  $-$69.6  &   12.9 &    7170.9927  &     96.3  &   33.3  \\
6714.0840  &    256.1  &   17.5 &    7171.2951  &     24.7  &   20.4  \\

\hline
\end{tabular}
\end{center}
\end{table}
%
%

\begin{table}
\begin{center}
\caption{RV measurements for 8 UMi from January 2010 to May 2015 using the BOES.}
\label{tab6}
\begin{tabular}{crr|crr}
\hline\hline JD         & $\Delta$RV  & $\pm \sigma$ &        JD & $\Delta$RV  & $\pm \sigma$  \\
$-$2450000 & m\,s$^{-1}$ &  m\,s$^{-1}$ & $-$2450000  & m\,s$^{-1}$ &  m\,s$^{-1}$  \\
\hline

5223.2931  &  $-$11.1  &   14.2 &   6739.0549  &     20.8  &   11.1  \\
5251.3004  &     52.3  &   10.1 &   6805.0734  &  $-$46.2  &   12.1  \\
5456.0690  &   $-$2.5  &   11.8 &   6823.0469  &     35.3  &   11.1  \\
5582.4015  &   $-$9.3  &   30.7 &   6972.3627  &  $-$36.4  &   12.5  \\
5671.1018  &  $-$29.0  &   10.6 &   7025.2594  &     89.2  &   21.2  \\
6026.2024  &   $-$2.2  &   12.7 &   7066.1797  &  $-$61.3  &   15.5  \\
6379.0953  &     63.6  &   10.9 &   7067.1929  &  $-$36.4  &   13.6  \\
6407.1638  &  $-$11.8  &    9.5 &   7068.1852  &  $-$60.4  &   13.3  \\
6437.2342  &  $-$20.6  &   11.6 &   7108.2615  &     60.6  &   18.4  \\
6679.2430  &   $-$1.2  &   10.0 &   7111.1635  &     31.5  &   17.4  \\
6710.2830  &  $-$25.9  &   10.7 &              &           &         \\

\hline
\end{tabular}
\end{center}
\end{table}

\end{document}